\documentclass[useAMS,usenatbib]{mnras}
\usepackage{amsmath}
\usepackage{float}
\usepackage{graphicx}
\usepackage{txfonts}
\usepackage{indentfirst}
\usepackage{ulem}
\usepackage[Symbolsmallscale]{upgreek}
\usepackage{nomencl}
\usepackage{enumitem}

\newcommand{\eq}[1]{\begin{equation}  #1 \end{equation}}

\newcommand{\br}[1]{\left( #1 \right)}

\newcommand{\bb}[1]{\left[ #1 \right]}

\newcommand{\dd}{{\rm d}}
\newcommand{\expo}[1]{~{\rm e}^{ #1 }}
\newcommand{\vek}[1]{\mbox{\boldmath $#1$}}

\newcommand{\svek}[1]{\mbox{\boldmath \scriptsize $#1$}}

\newcommand{\ic}{{\rm i}}

\def\araa{ARA\&A}
\def\apj{ApJ}
\def\apjl{ApJ}

\def\aap{A\&A}

\def\mnras{MNRAS}
\def\nat{Nature}

\def\prd{Phys.~Rev.~D}

\title[Acquiring the Lefschetz thimbles]{Acquiring the Lefschetz thimbles: efficient evaluation of the diffraction integral for lensing in wave optics}

\author[Shi]{Xun Shi$^{1}$\thanks{E-mail: xun@ynu.edu.cn} \\
$^{1}$South-Western Institute for Astronomy Research (SWIFAR), Yunnan University, 650500 Kunming, P. R. China}

\begin{document}
\maketitle
\begin{abstract} 
Evaluating the Kirchhoff-Fresnel diffraction integral is essential in studying wave effects in astrophysical lensing, but is often intractable because of the highly oscillatory integrand.
A recent breakthrough was made by exploiting the Picard-Lefschetz theory: 
the integral can be performed along the `Lefschetz thimbles' in the complex domain where the integrand is not oscillatory but rapidly converging.
The application of this method, however, has been limited by both the unfamiliar concepts involved and the low numerical efficiency of the method used to find the Lefschetz thimbles.  
In this paper, we give simple examples of the Lefschetz thimbles and define the `flow lines' that facilitate the understanding of the concepts.
Based on this, we propose new ways to obtain the Lefschetz thimbles with high numerical efficiency, which provide an effective tool for studying wave effects in astrophysical lensing.
\end{abstract}

\begin{keywords}
methods: numerical -- software: development -- gravitational lensing: strong -- gravitational lensing: micro
\end{keywords}

\section{Introduction}
Wave optics description of electromagnetic wave propagation after a phase screen is given by the Kirchhoff-Fresnel diffraction integral (e.g. \citealt{schneider92, born}). 
How to efficiently evaluate the Kirchhoff-Fresnel diffraction integral is thus the key problem in understanding diffraction and interference phenomena that arise, e.g., in the case of gravitational or plasma lensing of coherent radio sources such as pulsars \citep{hewish68}, fast radio bursts (FRBs, \citealt{lorimer07, thornton13}), and gravitational waves \citep{abbott16, agazie23, epta23,reardon23, xu23}.
Especially, scintillation and plasma lensing of radio pulsars and FRBs in the interstellar medium \citep[e.g.][]{rickett90, stinebring01, walker04, cordes06, brisken10, cordes17} often involve structures comparable or smaller than the Fresnel scale. 
Thus, a particular interest in evaluating the Kirchhoff-Fresnel diffraction integral exists in the efforts of understanding scintillation and plasma lensing caused by interstellar structures \citep{walker04, grillo18, jow20, jow21, shi21, jow22, jow23, shi24, shi24b}.  

The Kirchhoff-Fresnel diffraction integral is, however, non-trivial to calculate. 
Analytic solutions exist only for a few idealized situations (e.g. for point-mass lenses, see \citealt{deguchi86,jow20,leung23}). 
The eikonal approximation, i.e. approximating the wave field as discrete stationary phase points and computing the interference among them, is precise at the high-frequency limit but does not capture the full wave effects at finite frequencies. 
Due to the oscillatory nature of the integrand, numerical computation is in general challenging.
Fast Fourier Transform (FFT) is in principle applicable for computing the diffraction integral numerically for any lens shape.
However, its application is usually limited to the perturbative wave optics regime where neither the frequency nor the fluctuations in the phase screen are high, otherwise, an exceedingly large number of sampling grids would be necessary \citep[see e.g appendix A of][]{grillo18}.
Between the regimes of validity of the eikonal approximation and the perturbative wave optics, there is a regime of non-trivial wave effects \citep{shi24b}. 
No viable method exists for this regime until the recent introduction of the Picard-Lefschetz theory (see \citealt{witten10}) to radio astronomy by \citet{feldbrugge19,feldbrugge23}.

The Picard-Lefschetz theory  is an exact and versatile approach for dealing with multidimensional oscillatory integrals which has recently been applied to fields ranging from condensed matter systems to quantum cosmology \citep[e.g.][]{tanizaki16, feldbrugge17, mou19}.
Its essence is to use Cauchy's theorem to deform the interval of the integration into the complex plane. 
Adding the imaginary dimension gives the freedom to choose the integration contour on which the integrand is non-oscillatory and the integral converges the most rapidly.  
The resulting contours are called the ``Lefschetz thimbles", which are a sum of the steepest descent contours connecting the saddle points of the integrand generalized to the complex domain.  
Integration along these thimbles is computationally efficient and insensitive to numerical cutoffs, making it suitable for practical applications.

Despite this success, the Picard-Lefschetz method has not yet been widely used.
The main reason, apart from its conceptual difficulty, could be the limited numerical efficiency in finding the Lefschetz thimbles for a general lensing configuration.
Although performing the integration along the Lefschetz thimbles is numerically trivial, finding the Lefschetz thimbles is not. 
\citet{feldbrugge19} developed and advocated a numerical method called `flowing the integration domain'. 
It progressively distorts the integration domain into the complex plane using a differential equation.
This method has been adopted by almost all subsequent astrophysics works using the Picard-Lefschetz method \citep{feldbrugge20,feldbrugge23b,shi21, jow21,jow22, suvorov22, jow23,jow24}.
In particular, a numerical algorithm to realize this flow method for $N-$dimensional integrations was detailed in \citet{jow21}.   
However, as we shall demonstrate in this paper, an infinite number of flow steps is required for convergence to the Lefschetz thimbles for almost all points on the original integration domain.
In this `flow method', the entire Lefschetz thimbles are contributed by a tiny fraction of the original integration domain, namely the vicinity of a number of discrete points. 
Thus, a lot of refinement of these small intervals is necessary during the flow process, leading to a low numerical efficiency.
With this understood, the numerical method of finding the Lefschetz thimbles can be immediately improved upon.

The purpose of this paper is threefold: 
First, we would like to promote the familiarity with the Lefschetz thimbles by giving simple examples where their shapes are easy to understand.
These examples can also serve as test cases for verifying numerical convergence. 
Second, we would like to aid the understanding of Lefschetz thimbles by defining the `flow lines', 
which will help us show that each piece of Lefschetz thimble is flown from with an infinitesimal interval on the real axis.
Third, we describe ways to efficiently obtain the Lefschetz thimbles numerically.

\section{The diffraction integral and the Lefschetz Thimbles}
\label{sec:Lefschetz}
The wave amplitude received by an observer for a point source with a unit magnitude is given by the Kirchhoff-Fresnel diffraction integral  (e.g. \citealt{schneider92, born}, see also \citealt{kramer24} for a different approach)
\eq{
    E(\vek{\beta}) = \br{\frac{\nu}{2\uppi \ic}} \int \exp\bb{\ic \nu \phi} \dd^2 \vek{x} \,.
} 

This wave amplitude $E(\vek{\beta})$ is also referred to as the `transmission factor' of an intermediate phase screen acting as the lens \citep{schneider92}. 
It is contributed by all possible wave paths connecting the source, the observer, and a point $\vek{x}$ on the phase screen according to the Principle of Huygens.
Here, $\vek{\beta}$ is the location of the source line-of-sight on the phase screen. 
The coordinates $\vek{x}$ and $\vek{\beta}$ are made dimensionless by scaling with the size of the lens $a_{\rm lens}$ following the convention of \citet{shi21} and \citet{jow23}.  
The dimensionless parameter $\nu$ is addressed as the ``reduced frequency'' since it is proportional to the observing frequency. 
In a lensing system, it is given by
\eq{
    \nu  = \frac{2\pi (1+z_{\rm lens}) a_{\rm lens}^2}{\lambda \bar{D}} \,, 
}
where $\lambda$ is the wavelength, and $\bar{D} = D_{\rm l} D_{\rm ls} / D_{\rm s}$ is a combination of the distances to the lens $D_{\rm l}$, the source $D_{\rm s}$, and that between the source and lens $D_{\rm ls}$. 
On cosmological scales, these distances should be interpreted as angular diameter distances, and one needs to include the factor with the lens redshift $z_{\rm lens}$ to account for the redshift of the light.
From the above definition, one can derive that the reduced frequency is the square of the lens size - Fresnel scale ratio, $\nu = a_{\rm lens}^2 / r^2_{\rm F}$.
It is the key parameter that determines the importance of the wave effects \citep[see e.g.][]{shi24b}.



\begin{figure}
    \centering
    \includegraphics[width=0.42\textwidth]{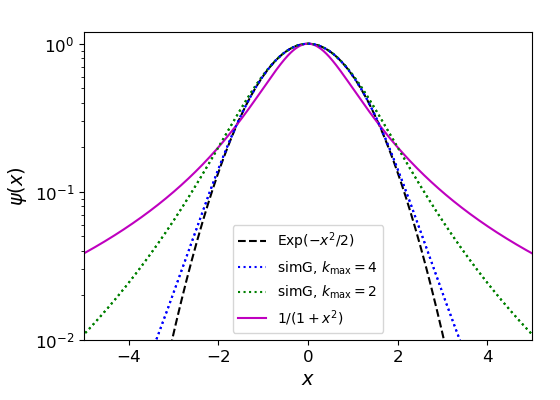}     
        \caption{
          Lens shapes used in this paper: a rational lens (magenta line) and a $k_{\rm max}=4$ simG lens (blue solid line).
          The Gaussian lens (black dashed line) and the $k_{\rm max}=2$ simG lens (green dotted line) are shown for comparison.
          As $k_{\rm max}$ increases, the simG lens rapidly approaches the Gaussian lens.
          }
        \label{fig:psi} 
\end{figure}

The phase function 
\eq{
    \phi(\vek{x}) = \frac{(\vek{x}-\vek{\beta})^2}{2} - \kappa\psi(\vek{x})
}
includes the geometrical phase delay and the dispersive phase delay from a lens potential with an amplitude $\kappa$ and a shape $\psi(x)$ (see Fig.\;\ref{fig:psi} for shapes used in this paper). 
We have adopted the convention that a converging lens (e.g. a gravitational lens) has a negative $\kappa$ value. 
Positive $\kappa$ values correspond to diverging lenses, which can represent overdense plasma lenses \citep[e.g.][]{clegg98, pen12, er14, cordes17, er18, er19, wagner20, shi21}.

From Eqs.\;1-3, one can see that how oscillatory the integrand is depends on the reduced frequency $\nu$ and the dimensionless lens amplitude $\kappa$.
For lensing systems with relatively large $\nu$ or $\kappa$ values e.g. plasma lensing of pulsars by the structures in the interstellar medium \citep[e.g.][]{hill05, brisken10, sprenger22}, the integrand is highly oscillatory and the diffraction integral is intractable with traditional numerical methods \citep{grillo18}.

The Picard-Lefschetz theory has been recently introduced to evaluate diffraction integrals in astrophysical lensing by \citet{feldbrugge19, feldbrugge23}.  
It provides an exact approach to computing highly oscillatory integrals for a wide range of lenses, 
and is highly complementary to traditional numerical methods such as FFT. 

By generalizing the integrand to the complex domain $\vek{x} \to \vek{z} = \vek{x} + \ic \vek{y}$ and separating the real and imaginary parts of the phase function,
\eq{
    \ic \phi \equiv h + \ic H \,,
}
the Picard-Lefschetz theory deforms the integration contour into a sum of the steepest descent contours of the $h$-function 
where the integrand is non-oscillatory and rapidly converging.
The resulting contours are called the `Lefschetz thimbles' -- a sum of smooth sub-contours connecting the saddle points of the integrand.

Each piece of Lefschetz thimble $\mathcal{J}_i$ is associated with one image $z_i$. 
By its definition, $\mathcal{J}_i$ is the steepest descent contours connected to the image $z_i$.
It is also all of the following:
\begin{enumerate}[label=(\roman*)]
    \item fix point of the downward flow of the $h$-function; 
    \item pathline of the downward flow starting from the image $z_i$;
    \item a subset (with descending $h$) of the constant phase contours connected to the image $z_i$. 
\end{enumerate}
See Fig.\;\ref{fig:ic3} for an illustration of the Lefschetz thimbles.

In addition to the Lefschetz thimbles associated with real saddle points i.e. the geometric images, there will generally be contributions from Lefschetz thimbles associated with saddle points with non-negligible imaginary parts i.e. the imaginary images \citep{jow21, shi24}, e.g., one marked by the black cross at Im(x)$\approx -1.5$ in Fig.\;\ref{fig:ic3}. 

\section{Lefschetz Thimble shape examples}
Here we show examples of the Lefschetz thimbles for a few simple cases.
For simplicity, we shall limit our discussion to 1D lenses in this and the next sections, and discuss a generalization to 2D situations in Section\;\ref{sec:2D}.

\subsection{Free propagation}

We first use the simplest case -- image in the absence of any lens to illustrate how an image is linked to the Lefschetz thimble.
In this case, the phase
\eq{
    \ic \phi = \ic \frac{(x + \ic y - \beta)^2}{2} = - (x-\beta) y + \ic \frac{(x-\beta)^2 - y^2}{2} \,.
}
There exists a unique real image at $x_0=\beta$, $y_0=0$, with $h_0 = H_0 = 0$. 

The Lefschetz thimble is the contour with constant imaginary phase $H=H_0$ and $h<h_0$ across the image. 
The single line satisfying these conditions is $y = x - \beta$ with slope $k=1$. 
Along this Lefschetz thimble, the integral becomes
\eq{
    E(\beta) = (1 + \ic ) \sqrt{\frac{\nu}{2\uppi \ic}} \int \exp \br{-\nu x^2} \dd x \,.
}
The integrand is a Gaussian distribution with width $\sigma_{\rm lens} / \sqrt{2 \nu} = r_{\rm F} / \sqrt{2}$.  
This matches the common wisdom that the Fresnel scale characterizes the size of the image in free propagation.

\subsection{In the vicinity of a lensed image}
Next, let us consider the case of a lens that creates additional lensed images.
In the existence of a lens, the shape of the Lefschetz thimbles still
asymptotes to $y=x-\beta$ far away from the lens at $x \to \pm \infty$.  

We now seek the expression for the slope $k$ of the thimble near an image in the complex plane $z_i$. 
Since the image is a stationary phase point with $\phi'(z_i) = 0$, the phase can be expanded around the image as
\eq{
    \phi \approx \phi(z_i) +  \phi''(z_i) \frac{\delta z^2}{2} \,.
}
Letting $\phi''(z_i) = \phi''_{\rm real, i} + \ic \phi''_{\rm imag, i}$, and writing $\delta z = (1 + \ic k) \delta x$, we get 
\eq{
    \delta h = \rm{Real}\br{\ic  \phi''(z_i) \frac{\delta z^2}{2}} = \bb{-\phi''_{\rm real, i} k  - \phi''_{\rm imag, i}\frac{1 - k^2}{2}} \delta x^2 \,,
}
\eq{
    \delta H = \rm{Imag}\br{\ic  \phi''(z_i) \frac{\delta z^2}{2}} =  \bb{\phi''_{\rm real, i}\frac{1 - k^2}{2} - \phi''_{\rm imag, i} k} \delta x^2 \,.
}
We can then obtain the value of $k$ from the condition that $\delta H = 0$ along a Lefschetz thimble.

For a real image, $\phi''_{\rm imag, i}=0$ and thus $k^2=1$. 
From the condition $\delta h = - \phi''_{\rm real, i} k \delta x^2 < 0$, one can derive $k=1$ for an image with the same parity with the source which has a magnification parameter $\mu_i \equiv 1/\phi''_{\rm real, i}>0$,
and $k=-1$ for an image with the opposite parity and $\mu_i < 0$.

For an imaginary image with $\phi''_{\rm imag, i}\ne 0$,
$\delta H = 0$ gives 
\eq{
    k = - \frac{\phi''_{\rm imag, i}}{\phi''_{\rm real, i}} \pm \sqrt{\br{\frac{\phi''_{\rm imag, i}}{\phi''_{\rm real, i}}}^2 + 1} \,,
}
i.e., the slope depends on the shape of the phase function around the image in the complex domain and has no bound.

\begin{figure}
    \centering
    \includegraphics[width=0.46\textwidth]{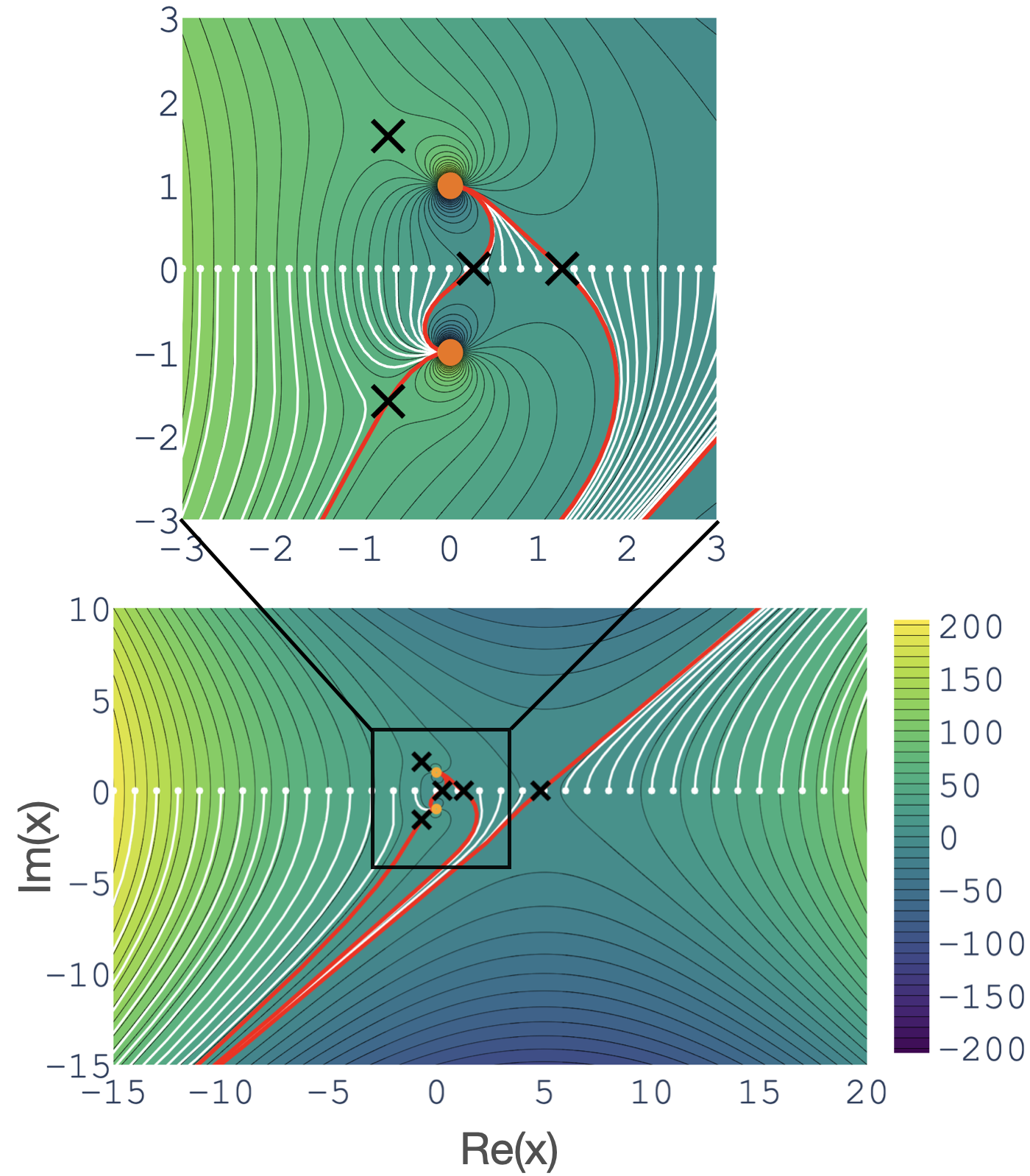} 
        \caption{
            \textbf{Flow lines} (white contours) for points on the real axis for a lensing system with a lens $\psi(x) = 1/(1+x^2)$ with an amplitude $\kappa=10$ placed at the origin, and the source at $\beta=5$. 
            The images are marked with black crosses, and the poles are marked with orange dots. 
            The Lefschetz thimble is shown as the red contour. 
            The contours of the $H$-function are shown in the background.
        } 
        \label{fig:ic3} 
\end{figure} 
\section{The flow lines}
The works of \citet{feldbrugge19, feldbrugge23} have established a link between the original integration contour -- the real axis and the contour of Lefschetz thimbles:  
the former can `flow' into the latter with a downward flow $\gamma_\ell$ of the $h$-function defined by
\eq{
\frac{\partial \gamma_\ell({z})}{\partial {\ell}} = -\nabla_{{z}} h[\gamma_\ell({z})] \,,
\label{eq:flow}
}
i.e., the flowed integration contour $X_{\ell}$ will converge to a set of steepest descent contours i.e. the Lefschetz thimbles as the number of flow steps $\ell \to \infty$,
\eq{
    \lim_{\ell \to \infty} X_{\ell} = \mathcal{J}\,.
    \label{eq:converge}
}

However, to consider the map between these two contours, one still needs to know how intervals on these contours map to each other.
For this purpose, we define `flow lines'  for points on the real axis, 
with a flow line tracking the trajectory of a point as it flows.
In other words, the flow line of a point is the pathline of the downward flow of the $h$-function starting from this point.

Fig.\;\ref{fig:ic3} presents the flow lines for points on the real axis for a lensing system with a lens of amplitude $\kappa=10$ at $\beta=5$.
The corresponding integral is $\int_{-\infty}^{\infty} \exp\bb{\ic \nu \bb{(x-5)^2/2 - 10/(1+x^2)}} \dd x$.
One can see that all flow lines end at either infinity or a pole,
and a piece of Lefschetz thimble $\mathcal{J}_i$ connects either an infinity and a pole, two poles, or two infinities.

The flow lines (white contours) help define the optimal generalized integration contour for an interval $(x_0, x_1)$ on the original integration axis.
When the flow lines associated with both $x_0$ and $x_1$ end at the same infinity, e.g. in the case of $x_0 = -10$ and $x_1 = -9$ in Fig.\;\ref{fig:ic3}, the integration over $(x_0, x_1)$ can be performed efficiently along the flow line of $x_0$ from $x_0$ to infinity and then again from infinity to $x_1$ along the flow line of $x_1$.

Remarkably, a whole piece of Lefschetz thimble $\mathcal{J}_i$ corresponds only to an infinitesimal interval on the real axis.
The location of this interval $a_i$ is where the real axis crosses the steepest ascending contour of the image.
For a finite interval that contains $a_i$, the optimal generalized integration contour then consists of $\mathcal{J}_i$ in addition to the flow lines of the two endpoints of the interval.
The flow lines for the special points $a_i$ are the most special. 
They flow first onto an image location $z_i$, and from there bifurcate into flows along the thimble piece $\mathcal{J}_i$ in two directions reaching infinities and/or poles.


\begin{figure*}
    \centering
    \includegraphics[width=0.6\textwidth]{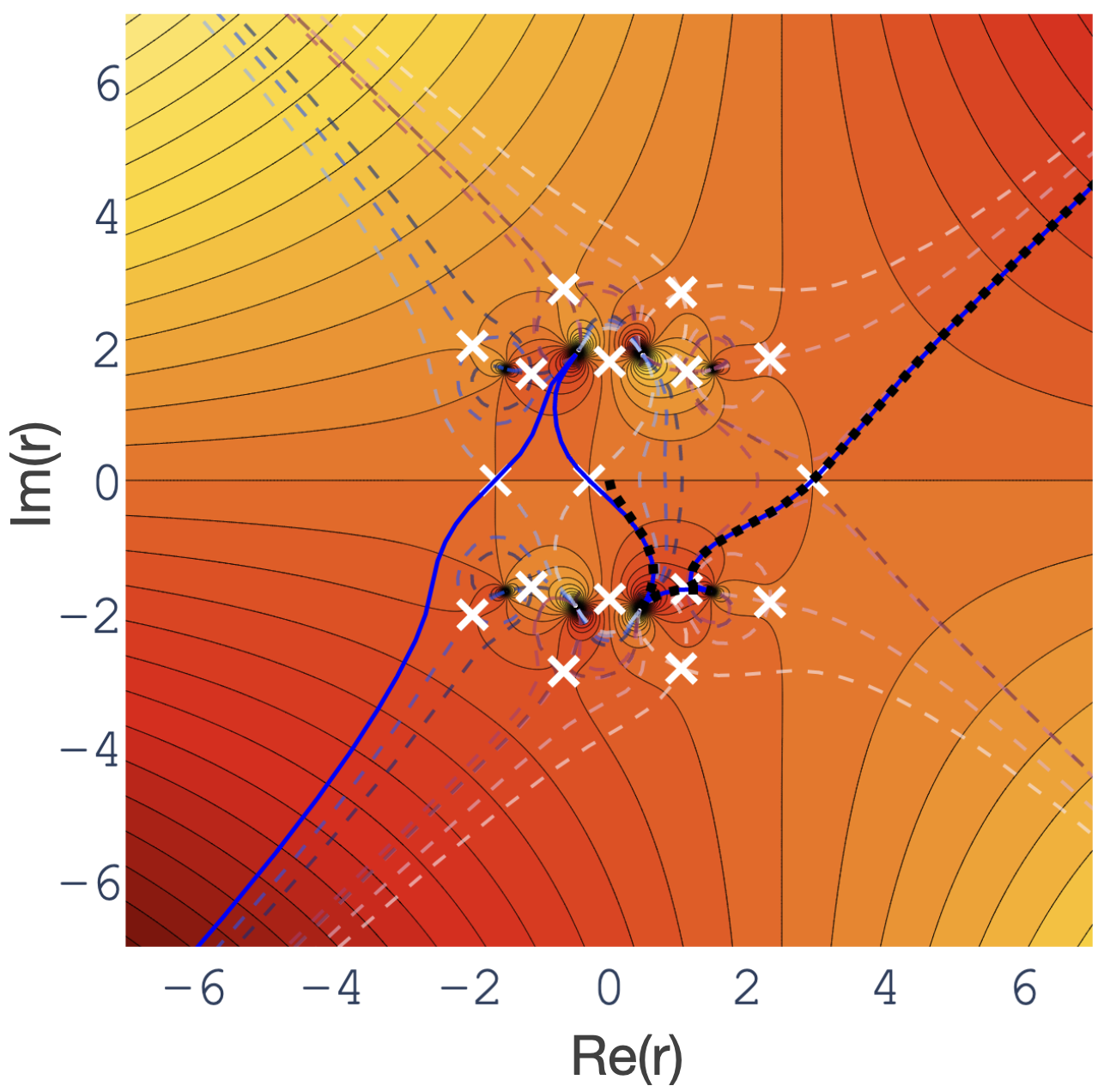} 
        \caption{An example of optimal integration contour found using the constant phase contour method for the radial integration of a 2D integration (see Eq.\;\ref{eq:2D}). 
        The source at $\beta_{x}=0$, $\beta_{y}=5$ is lensed by a converging 2D approximate Gaussian lens (see Fig.\;\ref{fig:psi}) with an amplitude $\kappa=-10$. 
        The polar coordinate of the 2D integration is taken to be $\theta = \uppi/6$ for this example.
        With the constant phase contour method, one first locates all stationary phase points (white crosses), 
        finds contours with $H_{\mathcal{J}_i} = H_i$ (dashed lines), and breaks them into steepest ascending and descending contours. 
        Then, one identifies the effective images from the fact that one of their associated steepest ascending contours crosses the real axis.
        Finally, one connects the steepest descending contours of the effective images to form the Lefschetz thimbles (the blue solid line).
        The optimal integration contour for the radial integration (black dotted line) connects the $r=0$ point to the Lefschetz thimble with the flow line for $r=0$.
        Contours of the $h$-function are shown in the background.
        }
        \label{fig:h_h8} 
\end{figure*}
\section{Finding the Lefschetz thimbles numerically}
That the Lefschetz thimble $\mathcal{J}_i$ is on the flow line of one point on the real axis has implications for the numerical method of finding $\mathcal{J}_i$.

The numerical method favored by \citet{feldbrugge19, feldbrugge23} and inherited by latter studies of the wave effects \citep{feldbrugge20,feldbrugge23b,shi21,suvorov22,jow21,jow22,jow23,jow24} is to flow the integrand using Eq.\;\ref{eq:flow}.
However, the flow lines demonstrate that the convergence of this flow to the Lefschetz thimbles (Eq.\;\ref{eq:converge}) occurs only at $\ell \to \infty$.
Most points sampled on the real axis will not contribute to the final Lefschetz thimble at all.
They only asymptote to $\mathcal{J}$ at $\pm\infty$ where $\mathcal{J}$ approaches the free propagation solution.
Numerically, to find a Lefschetz thimble piece $\mathcal{J}_i$ one needs to repeatedly refine intervals in the vicinity of the flow line of $a_i$. 
A lot of computation is spent on flowing the other intervals which are finally removed because their contributions turn out to fall below a threshold. 
This imposes a limitation on the numerical efficiency of this flow method in finding the Lefschetz thimbles.

With the current flow method, it is particularly difficult for the integration contour to converge to the Lefschetz thimbles in regions far away from the real axis in the complex domain.
However, the precise shape of the Lefschetz thimbles in these regions is increasingly important for the evaluation of the diffraction integral at smaller reduced frequency $\nu$, 
and such a shape is non-trivial for a lensing system with a lens well-separated from the source (i.e. with $|\beta| \gg 1$).
Most computations currently performed with the Picard-Lefschetz method avoid this problem by limiting to relatively high $\nu$ and small $|\beta|$ values (i.e. with small angular offsets between the source and the lens).
In such a regime, most line intervals quickly dip below a threshold in their contribution to the integral and are removed from the flow to improve numerical efficiency.
Although the small $|\beta|$, relatively high $\nu$ regime is relevant in many applications, it is not the only regime of interest.
In the study of pulsar scintillation, wave effects in the large $|\beta|$ regime are of particular interest. 
This regime is challenging for the current flow method, and the typical large lens amplitude $\kappa$ for pulsar scintillation \citep[e.g.][]{shi21b} makes the perturbative methods invalid already at small reduced frequencies \citep{jow23}.
A more efficient approach to finding the Lefschetz thimbles is particularly needed for this regime.

\subsection{The improved flow method}
To see how one can improve on this original method, let us revisit the different conceptual perspectives of a Lefschetz thimble piece $\mathcal{J}_i$ (see Sect.\;\ref{sec:Lefschetz}). 
The original flow method utilizes the first perspective (i), that $\mathcal{J}_i$ is the fixed point of the downward flow of the $h$-function. 
The second perspective (ii), that $\mathcal{J}_i$ is the pathline of the downward flow starting from $z_i$, gives a straightforward way to improve on it:
Instead of starting the flow from sampled points on the whole real axis, one just needs to start from points in the vicinity of each effective image. 
The pathlines i.e. flow trajectories of these points will yield the Lefschetz thimbles $\mathcal{J}$.
Clearly, this improved flow method can save a lot of computing power and memory compared to the original flow method.

One advantage of the original flow method is that it does not require an identification of the effective images.
The improved flow method, however, does require this identification. 
The identification can be achieved by first flowing points in the vicinity of all stationary phase points with the upward flow of the $h$-function to trace the ascending contours, and selecting the effective images as those with an ascending contour that crosses the real axis.
Considering this additional step, the improved flow method could still be more efficient than the original flow method for some lensing configurations.

\subsection{The constant phase contour method}
The third perspective (iii), that $\mathcal{J}_i$ is a subset (with descending $h$) of the constant phase contours connected to the image $z_i$, provides a completely different way to find the Lefschetz thimbles. 
Taking advantage of the fact that the phase function is non-oscillatory on a piece of Lefschetz thimble $\mathcal{J}_i$,
one can find $\mathcal{J}_i$ starting from globally mapping the constant phase contour $H_{\mathcal{J}_i} = H_i$. 
This results in a very efficient method to find the Lefschetz thimbles which we would like to promote in this paper and have applied to recent works \citep{shi24, shi24b}.
The algorithm is sketched below (see also a demonstration in Fig.\;\ref{fig:h_h8}):

\begin{enumerate}[label=(\arabic*)]
    \item get the saddle points of the phase function i.e. the images;
    \item get the relevant $H_i$ values at the images;
    \item get the contour $H_{\mathcal{J}_i} = H_i$ associated with each $H_i$ value;
    \item cut the contour at images, poles, and integration domain boundaries to obtain 2 descending (with $h<h_i$) + 2 ascending (with $h>h_i$) contour pieces linked to each image;
    \item select the effective images that are either real or with an ascending contour passing the real axis;
    \item connect descending contours of all effective images to form the overall Lefschetz thimbles $\mathcal{J}$.
\end{enumerate}

Unlike the flow method which is a dynamic method that utilizes predominantly the $h$-function, 
this constant phase contour method is a static method that relies on the global information of the $H$-function.
The idea of this method has been mentioned in \citet{feldbrugge19} and \citet{feldbrugge23}, but they thought it would have low numerical efficiency and disfavored it compared to the flow method.
However, given that fast contour-finding algorithms are available in most computational languages, the numerical efficiency of this method can be extremely high. 
With our sample script written in Python run on a laptop, it takes less than 2 seconds for the algorithm to locate the Lefschetz thimbles for the lensing configuration presented in Figure A2 of \citet{jow23}.
The numerical efficiency of this method is not sensitive to the range of integration intervals, and thus is particularly useful for lensing configurations with large source-lens angular offset $|\beta|$ and/or small reduced frequency $\nu$.
Another advantage of this new approach is that the precision of the resulting Lefschetz thimbles is set by the resolution of the grid for contour-finding, i.e. the precision is known and easily controllable.

Note that the constant phase domain has a dimension of $2N-1$ for an N-dimensional integral while the Lefschetz thimbles have a dimension of $N$. 
Their dimensions match only for one-dimensional integrals. 
We show how this method can be developed into an efficient method for a general 2D integral in the following.

\section{generalization to 2D integrals}
\label{sec:2D}
A great advantage of the original flow method is that it can be easily applied to oscillatory integrals of any dimension. 
The methods presented in this paper both require all stationary phase points in the generalized complex domain to be found as a first step.
This is numerically trivial for a 1D integral but is much harder for a higher dimensional integral with a general, asymmetric lens. 

However, as noticed by \citet{feldbrugge20} and \citet{tambalo23}, there is an easy way out for 2D integrals. 
When written in the polar coordinate, 
\eq{
    \label{eq:2D}
    E(\vek{\beta}) = \frac{\nu}{2\uppi\ic} \int_0^{2\uppi} \dd \theta \int_0^{\infty} r \expo{\ic\nu \phi(r,\theta,\svek{\beta})}\dd r \,,
}
the integral in $\theta$ is over a finite range and can be performed with standard numerical techniques\footnote{See \citet{feldbrugge23c} for caveats on the existence of path integrals.}.
Thus, the problem reduces to an evaluation of a 1D oscillatory integral in $r$. 
The only caveat is that the lower limit of integration is $0$ rather than $-\infty$.
One just needs to connect the Lefschetz thimbles with the flow line for $r=0$ to form a closed integration contour (see Fig.\;\ref{fig:h_h8}).

This solves the general problem of studying single-screen astrophysical lensing in the wave optics framework since 2D integrals provide an adequate description there.
We show an application of the constant phase contour method to compute the diffraction pattern produced by a 2D approximate Gaussian lens in the appendix.

Finally, we note the restriction of lens shapes that can be treated with the Picard-Lefschetz method. 
Since an analytic continuation into the complex plane is a key procedure in the Picard-Lefschetz method, the lens shape must be such that the phase function can be analytically continued into the complex plane.
For lensing by point masses, the phase involves a logarithm function which makes the generalization into the complex plane more intricate.
However, the logarithm potential represents a special case where the result of the Kirchhoff-Fresnel diffraction integral can be expressed analytically in terms of a special function \citep{nakamura99}.
The Gaussian lens shape $\psi \propto\expo{-x^2}$ also poses a difficulty by having an infinite number of saddle points in the complex plane and an essential singularity at infinity.
One can, nevertheless, approximate the Gaussian lens using a rational function $\psi(x) \propto 1/\sum_{k=0}^{k_{\rm max}} \frac{2^{-k} x^{2k}}{k!}$ to arbitrarily high precision (this lens shape is referred to as the `simG' lens, see Fig.\;\ref{fig:psi}).
Similar approximations can be made for any lens shape.

\section{Conclusion}
\label{sec:conclusion}
The recently introduced Picard-Lefschetz method has enabled the evaluation of the oscillatory Kirchhoff-Fresnel diffraction integral for a general lensing system. 
This paper is meant to clear the main obstacles to its application by clarifying the concept of the Lefschetz thimbles (the new integration contours in the complex domain) and proposing new efficient ways to find them.

We show that in the case of free propagation, the Lefschetz thimbles in the complex plane $\vek{z} = \vek{x} + \ic \vek{y}$ generated from each integration axis is a linear line with slope one: $\vek{y} = \vek{x}-\vek{\beta}$.   
In the presence of a lens, the Lefschetz thimbles are still asymptotic to this line far away from the lens.
In the vicinity of an image, the slope of the Lefschetz thimble is $k=\pm 1$ around real images and can have an arbitrary value determined by the shape of the phase function around imaginary images.

The `flow lines' are defined to show the pathlines of the downward flow of the $h$-function (real part of the phase $\ic \phi$).
With these pathlines, it is clear that the Lefschetz thimbles are mapped from tiny intervals on the real axis, each corresponding to an image.
Thus, the originally proposed `flowing the integration domain' method for finding the Lefschetz thimbles is far from optimal.
Although the flow method claims to let the integration domain converge to the Lefschetz thimbles, it requires an infinite number of flow steps and the convergence occurs only at $\pm \infty$ for almost all intervals on the original integration domain.
This renders this method numerically inefficient and makes it difficult to apply to lensing at very small reduced frequencies and/or lensing systems with large angular separations between the lens and the source.

We propose new methods of finding the Lefschetz thimbles that either start the flow from points in the vicinity of the images or use the constant phase contour method.
The latter method which is based on the global information of the $H$-function (complex part of $\ic \phi$), is the prime method for 1D integrals and is particularly efficient and accurate.
We have sketched the algorithm for this method and shown that it can also be developed into an efficient method for 2D integrals.

With these insights into the Lefschetz thimbles and the efficient ways to locate them, the oscillatory nature no longer poses a challenge to the integration of the Kirchhoff-Fresnel diffraction integral.
This provides an effective tool for the study of wave effects for a wide range of lensing configurations.

\section*{Data Availability Statements}
\noindent No new data were generated or analyzed in support of this research.

\section*{Acknowledgements}
\noindent X.S. thanks the referees for the helpful comments and suggestions. This work is supported by NSFC No. 12373025.


\bibliographystyle{mnras}

\appendix

\begin{figure*}
    \centering
    \includegraphics[width=0.98\textwidth]{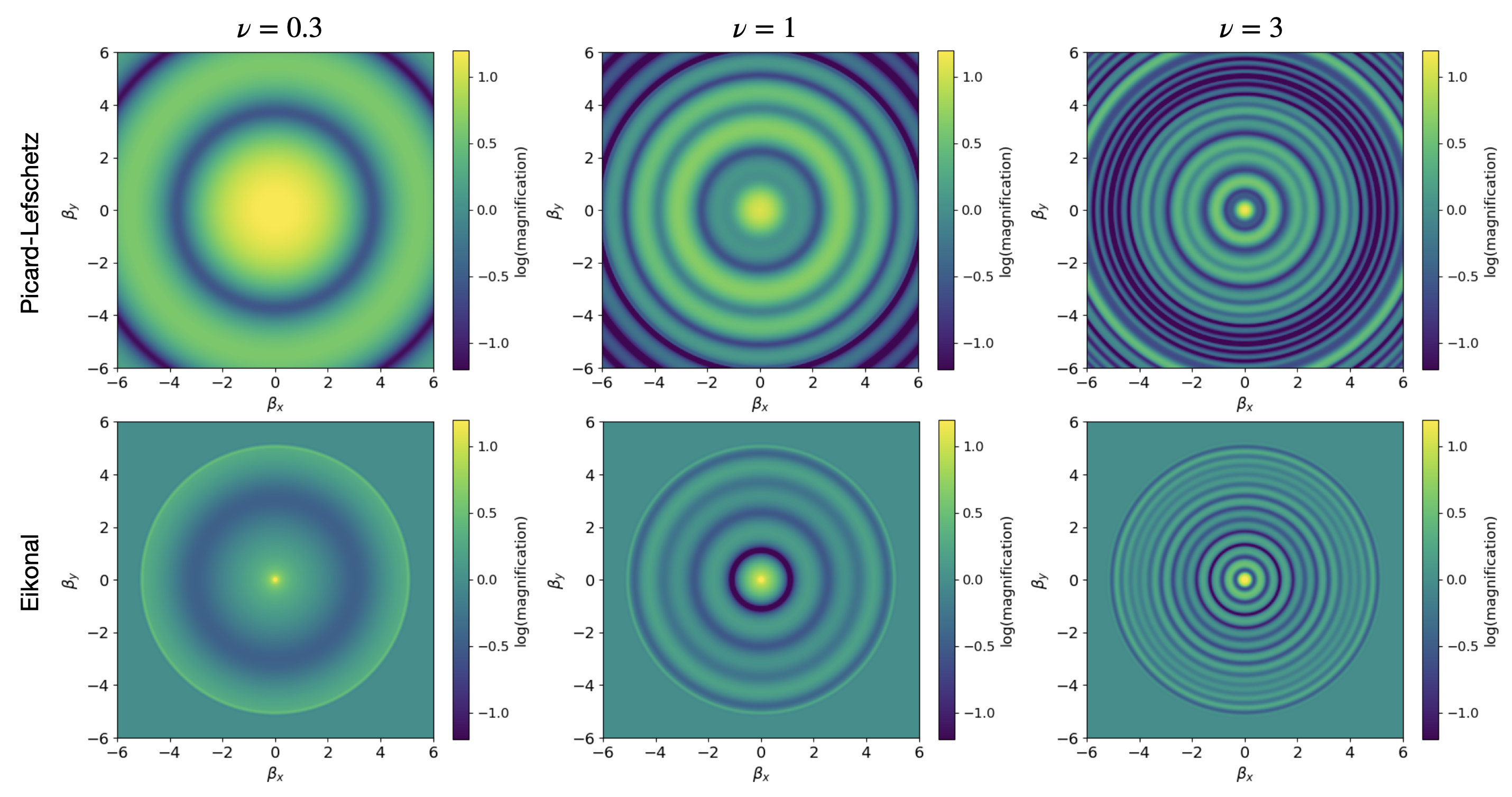}     
        \caption{Diffraction patterns computed using the numerical method outlined in Section\;\ref{sec:2D} (upper panels).
        Shown are the magnifications of a point source ($|E|^2$) at various source locations $\beta_x$ and $\beta_y$ after being lensed by a 2D approximate Gaussian lens (see Fig.\;\ref{fig:psi}) with an amplitude $\kappa=-10$.
        As a comparison, the diffraction patterns computed using the eikonal approximation are shown in the lower panels. 
        The reduced frequency $\nu$ is set to $0.3$, $1$, and $3$ for the columns rows, respectively.
        } 
        \label{fig:2D} 
\end{figure*}

\section{Diffraction by a 2D lens}
As an application of the constant phase contour method to the evaluation of the Kirchhoff-Fresnel diffraction integral, we compute the diffraction pattern produced by a 2D lens.
The lens is an axisymmetric, converging lens with an approximate Gaussian (`simG') shape with $k_{\rm max} = 4$ (see Fig.\;\ref{fig:psi}) and an amplitude $\kappa=-10$.

In Fig.\;\ref{fig:2D} we show the diffraction pattern as a function of the 2D source location $\beta_x$ and $\beta_y$ for three different reduced frequencies.
The upper panels are results computed with the Picard-Lefschetz method using the numerical approach outlined in Section\;\ref{sec:2D}.  
The lower panels are results computed with the eikonal approximation. 
The chosen frequencies are within the range where wave effects are non-trivial \citep{shi24b}. 
In this range, the results considering full wave effects differ significantly both from their eikonal approximations and also from the unlensed cases.
Notably, there exists a caustic at $|\vek{\beta}| = 5.1$ outside which there exists only one real image, and thus the eikonal results show no flux variation outside the caustic.
Within the caustic, the flux under the eikonal approximation is computed from the interference of the three real images created by the lens.
The Picard-Lefschetz method, on the other hand, captures the full wave effects and shows a clear diffraction pattern both inside and outside the caustic that differs from the eikonal results. 

The numerical speed for finding the Lefschetz thimbles with the constant phase contour method depends on the complexity of the lens.
The required time increases with the number of stationary phase points.
For the $k_{\rm max} = 4$ simG lens with 17 stationary phase points (Fig.\;\ref{fig:h_h8}), our sample code takes $\sim$10 seconds for each radial integral of Eq.\;\ref{eq:2D} using one core. 
The required time for each pixel in the diffraction pattern also depends on the sampling of the angular integral.
The necessary $\theta$ sampling heavily depends on the reduced frequency $\nu$. 
For a higher $\nu$, the integrand over $\theta$ is more oscillatory.
We have used 120 $\theta$-points for $\nu = 3$.
This is close to the maximum number of $\theta$-points needed since for $\nu \gtrsim 10$ the eikonal approximation would work well and there is little need for full integration of the Kirchhoff-Fresnel diffraction integral.

\end{document}